\documentclass[twocolumn,pra,aps,nofootinbib,superscriptaddress,amsmath,amssymb,showpacs]{revtex4}
\usepackage{graphicx,color}
\usepackage{epstopdf}

\newcommand{\fig}[1]{Fig.~\ref{#1}}
\newcommand{\beq}{\begin{equation}}
\newcommand{\eeq}{\end{equation}}
\newcommand{\bqa}{\begin{eqnarray}}
\newcommand{\eqa}{\end{eqnarray}}
\newcommand{\nn}{\nonumber}
\newcommand{\nl}{\nn \\ &&}
\newcommand{\dg}{^\dagger}

\newcommand{\erf}[1]{Eq.~(\ref{#1})}
\newcommand{\erfs}[1]{Eqs.~(\ref{#1})}
\newcommand{\eqn}[1]{\erf{#1}}

\newcommand{\ket}[1]{\left|{#1}\right\rangle}

\newcommand{\ito}{It\^o }

\newcommand{\cu}[1]{\left\{ {#1} \right\}}

\newcommand{\D}{\mathcal D}
\newcommand{\hsys}{H_\mathrm{sys}}
\newcommand{\hc}{\mathsf{H.c.}}
\DeclareMathOperator{\sech}{sech}

\newcommand{\infid}{\bar{\mathcal{F}}}
\newcommand{\mbs}{$M$-BS}
\newcommand{\source}{\textsf{s}}
\newcommand{\target}{\textsf{t}}
\newcommand{\fic}{\textsf{f}}

\begin{document}

\title{An approximate method for treating dispersion in one-way quantum channels}

\author{T. M. Stace}
\address{DAMTP, University of Cambridge, CB30WA, UK}
\author{H. M. Wiseman}
\address{Centre for Quantum Computer Technology, Center for Quantum Dynamics, School of Science, Griffith University, Nathan 4111, Australia}


\begin{abstract}
Coupling the output of a source quantum system into a target quantum system is easily treated by cascaded systems theory if the intervening quantum channel is dispersionless. However, dispersion may be important in some transfer protocols, especially in solid-state systems. In this paper we show how to generalize cascaded systems theory to treat such dispersion, provided it is not too strong. We  show that the technique also works for fermionic systems with a low flux, and can be extended to treat fermionic systems with large flux.  To test our theory, we calculate the effect of dispersion on the fidelity of a simple protocol of quantum state transfer. We find good agreement with an approximate analytical theory that had been previously developed for this example.
\end{abstract} 

\maketitle

\section{Introduction}

Theoretical methods for treating non-ideal components in quantum networks is an important task for quantifying imperfections in experiments.  One common example  is photon loss in optical channels, which can be treated by invoking a fictitious beam splitter that mixes the channel mode with other experimentally inaccessible modes \cite{gar00}.   The component of the channel mode reflected by the beam splitter therefore corresponds to photon loss. This approach also allows inefficient detection to be accurately modeled.

In this paper, we present a technique for treating dispersion in quantum channels.  Dispersion arises when modes acquire a phase after propagation that depends non-linearly on frequency.  Typically, efforts are made to operate optical fibres at the zero-dispersion point in order that this effect be small, and heterogeneous structures may be used to provide an effectively dispersion free channel.  Nevertheless, in some circumstances, it may be desirable to operate in a regime where dispersion is not negligible.  Recent proposal for implementing mesoscopic analogues of optical schemes, such interferometers \cite{ji03,Chung05}, and quantum state transfer protocol \cite{stace:126804}  using the quantum Hall effect will necessarily have some dispersion, due to the non-zero mass of quasi-electrons in the edge state.  In that case, an \emph{ad hoc} approach was used to estimate the effect of dispersion. Another quantum systems in which dispersion during propagation is expected to be important is atom lasers \cite{Wis97}. 

In the example of treating photon loss, an additional element, the beam-splitter, is added to an otherwise ideal channel to provide a tractable model.  In analogy with this approach, we also introduce an extra element to an otherwise ideal (i.e.\ dispersionless) channel: a resonant, damped cavity operating in reflection.  Near resonance, incident modes suffer a  frequency dependent phase shift on reflection, depending non-linearly on their detuning from the resonance.  This is broadly the same condition that arises in a dispersing channel, so the aim is to fix the resonance and damping of the cavity to match dispersion as closely as possible.

Since there are only two parameters for the cavity, it is plainly not possible to treat arbitrary dispersion with this approach.  However, we show that in simple networks (without feedback or interference between different paths) it is possible to match up to third order in the dispersion relations. Thus our approach handles channels that are not-too-dispersive, over the range of input frequencies.

We begin by summarising the effects of both dispersion and reflection from a cavity.  We then derive  the conditions for which cavity reflection is a good approximation to a dispersive channel, relating the frequency and damping of the fictitious cavity to the physical parameters describing the dispersive channel.  We then make some brief comments on the restrictions of this approach to channels in feedback systems, and fermionic systems, and derive a master equation for describing the dynamics for subsystems connected by a one-way quantum channel.  The paper ends with a simple example  illustrating the application of the approach to treating quantum state transfer over weakly dispersive channels.

\section{Preliminaries}

Consider the case of noninteracting quantum field propagating in one dimension. Let $\hbar= 1$. Then at the origin (e.g. point of emission) the field can be expanded in terms of eigen-mode operators 
\beq
\psi(0) = \sum_\omega b_\omega e^{-i\omega t}.
\eeq
Here we are implicitly considering only modes propagating in the positive direction. This limitation will be justified by later (more restrictive) assumptions. The use of a discrete sum is for notational convenience only. A widely applicable expression for the  dispersion relation is  
\beq
\omega = vk + \alpha k^2.
\eeq
The group velocity is 
\beq
u = \left.\frac{\partial \omega}{\partial k}\right|_{\omega = \bar{\omega}} = \sqrt{v^2 + 4\alpha \bar{\omega}}, 
\eeq
where $\bar{\omega}$ is the carrier frequency. 
For a free nonrelativistic particle $v=0$ and $\alpha = 1/2m$. For an electron propagating in an edge state typically $\alpha \bar\omega \ll v^2$ so that $u \approx v$ \cite{stace:126804}. (We will return later to the problem that an electron is not a boson.) 
At position $L$ the field is
\beq \label{psiL}
\psi(L) = \sum_\omega b_\omega e^{-i\omega t + i k(\omega)L},
\eeq
where
\beq
k(\omega) = (2\alpha)^{-1}(-v+\sqrt{v^2 + 4\alpha \omega}).
\eeq

Now compare the above expressions to a dispersionless boson field. At the origin we again have
\beq
\phi(0) = \sum_\omega b_\omega e^{-i\omega t}.
\eeq
The (non)-dispersion relation is $\omega = c k$, so at position $l$ the field is
\beq
\phi(l) = \sum_\omega b_\omega e^{-i\omega t + i\omega l/c}.
\eeq
If however we also include (a) a global phase shift and (b) bouncing off a single-mode cavity of central frequency $\omega_\fic$ and linewidth $\gamma_\fic$ then
\beq  
\phi(l) = \sum_\omega b_\omega e^{-i\omega t + i\omega l/c + i\theta} \frac{\gamma_\fic + 
2i(\omega-\omega_\fic)}{\gamma_\fic - 2i(\omega-\omega_\fic)} .\label{filter}
\eeq
For this result, see for example Ref.~\cite{wal94}. 
This is valid only if the Markovian description of the coupling of the external field to a single mode can be used, which requires 
\beq \label{forMarkovian}
\Delta_\fic, \gamma_\fic , \delta\omega \ll \bar{\omega},
\eeq
where $\Delta_\fic = \bar\omega - \omega_\fic$ and $\delta\omega$ is the uncertainty in the energy. 

\section{Feedforward}

Consider the case where the output of system $\source$ (source) is the input to system $\target$ (target). To model dispersion in the propagation between $\source$ and $\target$ we consider an non-dispersing reflecting off an intermediate (fictitious) cavity mode $c_\fic$, as shown in \fig{fig:TripleCascade}. From \erfs{psiL} and (\ref{filter}), this will work if we can make the approximation 
\beq
\frac{(-v+\sqrt{v^2 + 4\alpha \omega}) L}{2\alpha} \approx \frac{\omega l}{c} + \theta + 2\arctan\frac{2(\omega-\omega_\fic)}{\gamma_\fic} \label{matchphases}
\eeq
In this feedforward case the   time delay $l/c$ in the propagation, and the absolute phase of the field $\theta$, are irrelevant to how system $\target$ responds to the output of system $\source$, as long as any classial driving fields have their timings and phases adjusted appropriately. Thus we can always choose $l$ and $\theta$ so that the constant and linear term in the expansion of the LHS of \erf{matchphases} about $\bar\omega$ agree with the RHS. Thus in choosing $\gamma_\fic$ and $\omega_\fic$ we need consider only higher order derivatives. Since we have two free parameters it is natural to look at the second and third derivatives.  Equating second and third derivative gives
\begin{eqnarray} 
\alpha  L/u^{3}& =& 16\gamma_\fic\Delta_\fic/(\gamma_\fic^2+4\Delta_\fic^2)^2\label{second_der}\\
6\alpha^2 L/u^{5}&=&16\gamma_\fic(12\Delta_\fic^2-\gamma_\fic^2)/(\gamma_\fic^2+4\Delta_\fic^2)^3 
\end{eqnarray} 
Solving for $\Delta_\fic$ and $\gamma_\fic$ yields
\begin{eqnarray}
\gamma_\fic^2 &=& 12\Delta_\fic^2(1+O(\sqrt{{{\alpha}/{L u}}})),\label{gamma}\\
\Delta_\fic^2 &=&\frac{\sqrt{3}u^3}{8L\alpha}(1+O(\sqrt{{{\alpha}/{L u}}})).\label{Delta}
\end{eqnarray}
The error is small when $\alpha\ll L u$. This is equivalent to $\tau_p \ll \tau_d$, where $\tau_p=L/u$ is the propagation time,  and $\tau_d=L^2/\alpha$ is the time for a pulse to disperse over a length scale $\sim L$.

In the weak dispersion limit of $v^2 \gg \alpha\bar{\omega}$, we have $\Delta_\fic^2/\bar\omega^2 = 
O(v^3/\alpha\bar\omega^2 L) = 
O( v / \bar\omega L) O( v^2 /\alpha \bar{\omega})  \gg O( v / \bar\omega L) = O(1/\bar{k}L)$. Thus from \erf{forMarkovian} we have 
\beq \label{minpropdist}
\bar{k}L \gg 1.
\eeq
In the opposite limit of $v^2 \ll \alpha\bar{\omega}$, we have $\Delta_\fic^2/\bar\omega^2 = 
O( \sqrt{\alpha/\bar\omega}/ L) = O(1/\bar{k}L)$. Thus \erf{minpropdist} applies in all regimes. 
 It might seem surprising that our description puts a {\em lower} limit on the propagation distance, that it be much longer than a mean wavelength. This can be understood as follows. If dispersion were significant (such that it is necessary to match up to the third derivative in \erf{matchphases}) over the  distance of a wavelength, the problem would be so non-Markovian that the cavity description would necessarily fail. If it is deemed necessary only to match up to the second derivative then in principle \erf{minpropdist} need not hold. However on physical grounds the second system cannot be within a wavelength or so of the first without a break-down of cascaded systems theory altogether. 
 Another consideration on the limitation of validity of the theory is that for the third order expansion to be a good approximation we must have 
\beq \label{lov}
\delta \omega \alt \gamma_\fic.
\eeq
This puts an {\em upper} bound of $L$ which scales as $(\delta \omega)^{-2}$. 

If all of the above conditions hold then we can write down a master equation for the cascaded systems $\source$, $c$ and $\target$ that will be a good description of dispersive propagation from $\source$ to $\target$. 

\section{Feedback or Interference}

In other situations the absolute time delay {\em does} matter, in particular with feedback. That is, if $\source$ feeds into $\target$ which feeds back into $\source$. In that case if we wish to use the master equation description we cannot include a time delay $l/c$. Thus the {\em first} derivative term must come from the cavity. This gives
\beq
(v^2+4\alpha\bar\omega)^{-1/2} L = \frac{4\gamma_\fic}{(\gamma_\fic^2+4\Delta_\fic^2)} 
\eeq
Substituting this into \erf{second_der} gives
\beq
\alpha (\gamma_\fic^2+4\Delta_\fic^2) = 2\Delta_\fic (v^2+4\alpha\bar\omega)
\eeq
From \erf{forMarkovian} we see that we have an inconsistency. Thus we cannot describe feedback for a dispersive field using this model. On the other hand, if $\alpha=0$ (no dispersion) then we can validly satisfy these equations with $\Delta_\fic = 0$ and $\gamma_\fic = 4 v/L$. Interestingly, \erf{forMarkovian} again gives \erf{minpropdist}. 

Another situation where time delays matter, at least the difference between two time delays, is when there are two paths by which system $\source$ may affect system $\target$. In that case, if the time difference is comparable to the total propagation time then the same inconsistency as noted above will arise. Thus the applicability of this approach to modelling dispersion is most promising for  
a simple forward chain, and we concentrate on this for the remainder of this paper.

\section{Master Equation}

\begin{figure}
\includegraphics[width=6cm]{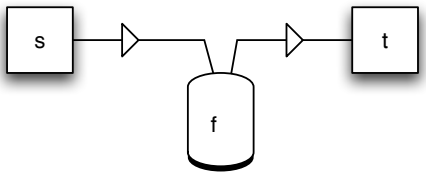}
\caption{Schematic of a triply cascaded system.  The output of subsystem $\source$ reflects off subsystem $\fic$, and the reflected field drives subsystem $\target$. No signal propagates in reverse.   
}
\label{fig:TripleCascade}
\end{figure}

To begin the quantitative analysis, we derive a general master equation for a triply cascaded system, shown in \fig{fig:TripleCascade}, where the outer systems are arbitrary, but the subsystem $\fic$ plays the role of the fictitious cavity introduced to simulate dispersion.  We assume that subsystem $i \in \cu{\source,\fic,\target}$ is linearly coupled to the external modes, $b_\omega$ according to 
\beq
H_{i-\textrm{coup}}=\sum_\omega \kappa_{i\omega} c_i b_\omega^\dagger+\kappa_{i\omega}^*  b_\omega c_i^\dagger.
\eeq
We compute the Heisenberg equation of motion for an arbitrary operator, $o_i$ of subsystem $i$, and make the Born-Markov approximation, in which we assume $\kappa_{i\omega}=\sqrt{\gamma_i/2\pi}$ is independent of $\omega$.  The resulting equation is a Stratonovich SDE.  In order to derive a master equation, we convert this into an \ito equation, taking care of the spatial ordering of the three cavities (see for example Ref.~\cite{WisemanThesis}). Alternatively, we can directly apply the cascaded systems theory of Refs.~\cite{Car93a,Gar93}, iterating the result to include the third system. 
The master equation for the state matrix for the triply cascaded quantum system is
\beq
\dot \rho=-i[\hsys + \tilde{H},\rho]+\D[\sqrt{\gamma_\source}c_\source+\sqrt{\gamma_\fic}c_\fic+\sqrt{\gamma_\target}c_\target]\rho ,
\label{cascadeME}
\eeq
where
\beq
\tilde H = \frac{i}{2}(\sqrt{\gamma_\source\gamma_\fic}c_\source^\dagger c_\fic+
\sqrt{\gamma_\fic\gamma_\target}c_\fic^\dagger c_\target+
\sqrt{\gamma_\target\gamma_\source}c_\source^\dagger c_\target-\hc) \nonumber
\eeq
and we have introduced the Lindblad superoperator  $\D[a]\rho=a\rho a^\dagger-(a^\dagger a \rho+\rho a^\dagger a )/2$. This master equation satisfies the requirement that dynamics in subsystem $\source$ is unaffected by the dynamics of subsystems $\fic$ or $\target$, and subsystem $\fic$ is unaffected by subsystem $\target$, as implied by the cascaded description.  We have also defined the bare Hamiltonian for the uncoupled systems $\hsys=H_\source+H_\fic+H_\target$.  $H_\source$ and $H_\target$ can be arbitrary, depending on the particular application in mind.  The middle subsystem is the fictitious cavity that serves to model dispersion, so we take  $H_\fic=\omega_\fic c^\dagger_\fic c_\fic$.

\section{Fermions}

\begin{figure}
\includegraphics[width=6cm]{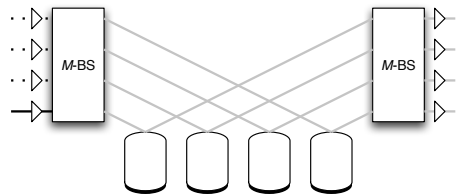}
\caption{Fermionic dispersion treated using $M$-port beam splitters to direct modes onto separate cavities, which are subsequently recombined. Dotted lines represent unoccupied modes, and grey lines indicate weakly occupied modes.
}
\label{fig:MBS}
\end{figure}

The technique described above was formulated for bosons. Where it breaks down for fermions is that  the Pauli exclusion principle permits only a single particle per cavity mode, so that the simple linear transformation resulting from reflection off a single cavity mode (\ref{filter}) does not hold. However, 
if there is at most one fermion involved in the problem, then particle statistics are irrelevant and our approach can be applied. Even  if there are many fermions, if the flux is low enough then our approach is applicable. Specifically, for a fermion flux of $n$ per second, the average occupation of the fictitious cavity is at most $N=n/\gamma_\fic$, so the proposal is restricted to fluxes $n\ll\gamma_\fic$.
That is, $n \ll \sqrt{u^3/L\alpha}$. 

One method to extend the regime of validity of our method in fermionic systems is shown in \fig{fig:MBS}.
  Here the output from $\source$, plus $M-1$ modes in the vacuum state, are directed through an $M$-port beam splitter (\mbs) onto $M$ fictitious cavities.  In this case, the average number of fermions, $N$, is distributed over $M$ cavities, so the mean occupation per cavity is $N/M$, which can be made small for sufficiently large $M$.  The splitting is then reversed, and the  $M$ modes drive the final subsystem $\target$. Physically, it is easiest to imagine that the output of $\source$ is a radially symmetric mode, and that the additional $M-1$ vacuum modes are being higher-order transverse modes. The fictitious \mbs\  then could simply be a device that separates $M$ transverse segments (e.g.\ wedges of a circular wire) and sends them to $M$ fictitious cavities. 
  
  The procedure just described leads to the following master equation:
  \bqa
\dot \rho &=& -i[\hsys + \tilde{H},\rho] \nl{+} \frac{1}{M} \sum_{k=1}^{M}  \D[\sqrt{\gamma_\source}c_\source+\sqrt{\gamma_\fic}c_k+\sqrt{\gamma_\target}c_\target] \rho,
\eqa
where 
\beq
\tilde H = \frac{i}{2M} \sum_{k=1}^{M}   (\sqrt{\gamma_\source\gamma_\fic}c_\source^\dagger c_k+
\sqrt{\gamma_\fic\gamma_\target}c_k^\dagger c_\target+
\sqrt{\gamma_\target\gamma_\source}c_\source^\dagger c_\target-\hc) \nonumber
\eeq
Here $\hsys$ is as before, but with $H_\fic = \sum_{k=1}^{M} \omega_\fic c\dg_k c_k$. 
It might be thought that a simulation with so many systems would be computationally expensive, but since it is only valid if each fictitious cavity has at most one excitation anyway, the Hilbert space dimension of the fictitious system as a whole is only $2^M$. Moreover, the probability that many 
[that is, $O(M)$] of the cavities are occupied at any one time is very small (since the occupation probability $N/M$ for any one cavity is assumed small). Thus, it should be possible to reduce the number of basis states required for a simulation dramatically. 


\section{Example: Quantum State Transfer}

In order to demonstrate our method, we apply it to a proposed scheme for quantum state transfer \cite{cir97} between two remote atoms each in a separate cavity, which are connected by an optical channel.  This scheme has been adapted to mesoscopic systems, using quantum dots instead of atoms and cavities, and quantum Hall edge states as a communication channel \cite{stace:126804}, so is relevant to both atom-optical and solid-state systems. This system was sufficiently simple that it was possible to find an approximate analytical expression for the effect of dispersion  \cite{stace:126804}.
Here we compare this approximation with the more sophisticated  method we have developed here.  

The protocol works by controlling the coupling strength between the atom and the cavity, $\Omega_{\source,\target}(t)$, at each site in such a way that the evolution coherently maps excitation in one atom to excitation in the other atom.  For an ideal channel, one class of suitable control pulses satisfies the relation $\Omega_{\source}(
t)=\Omega_{\target}(\tau_p-t)=\Omega(t)$.   Dispersion in the intervening channel has two effects on the fidelity of the transfer protocol.  Firstly, the dispersion will broaden the wavepacket in the channel so that it will have some reduced fidelity with respect to a comparable wavepacket in an ideal, dispersionless channel.  Secondly, dispersion modifies the group velocity slightly, so that the wavepacket arrives at the destination at a slightly different time.  This can be accounted for simply by adjusting the timing and phase of the control fields so that the term linear in $\omega-\bar\omega$ in the expansion of \eqn{matchphases}, is zero, i.e.\ $\tau_p=l/c+4\gamma/(\gamma^2+4\Delta^2)\approx l/c+\sqrt{3}/2\Delta$. 
For the purposes of feed-forward simulation, we can take $\tau_p=l/c=0$, so the conditions on the driving fields for optimal transfer is $\Omega_{\source}(
t)=\Omega_{\target}(\sqrt{3}/2\Delta-t)$.

For this model we consider $H_{i=\source,\target}=\omega_i (c^\dagger_i c_i+a^\dagger_i a_i)+\Omega_i(t)(c_i^\dagger a_i+a_i^\dagger c_i)$, where $c_i$ are cavity mode annihilation operators, and $a_i$ are atomic lowering operators for each subsystem $i$, and $\Omega_i(t)$ is a controllable coupling between the atom and cavity mode.   We assume the ideal case, $\omega_i=\bar \omega$ and $\gamma_{\source,\target}=\bar \gamma$.  Moving to the usual interaction frame, the system Hamiltonian is
\beq
\hsys=\sum_{i=\source,\target}\Omega_i(t)(c_i^\dagger a_i+a_i^\dagger c_i)-\Delta c_2^\dagger c_2.
\eeq
We assume the system starts in the state $\ket{e,0;0;g,0}$, where $\ket{\mathrm{atom_\source},\mathrm{cavity_\source};\mathrm{cavity_\fic};\mathrm{atom_\target},\mathrm{cavity_\target}}$ denotes the states of the three subsystems expressed in the energy eigenbasis of the atoms and cavities. Because there is at most excitation, this system is equivalent to a fermion system \cite{stace:126804}, and there is no need for more than one fictitious cavity. 

  We can now solve \eqn{cascadeME} for the state matrix of the system, which is spanned by the states
\begin{eqnarray}
\{\ket{g,0;0;g,0}, \ket{e,0;0;g,0}, \ket{g,1;0;g,0},\nonumber\\
\ket{g,0;1;g,0}, \ket{g,0;0;g,1}, \ket{g,0;0;e,0}\},\nonumber
\end{eqnarray}
We use a simple pulse sequence that implements state transfer  $\Omega_{\source,\target}(t)=\bar \gamma \sech(\bar \gamma t/2)/2$ \cite{sta02}. Recall that we are using the standard convention for cascaded systems that the origin of time for system $\target$ is delayed with respect to that for system $\source$. 

\begin{figure}[t]
\begin{center}
\includegraphics[width=8cm]{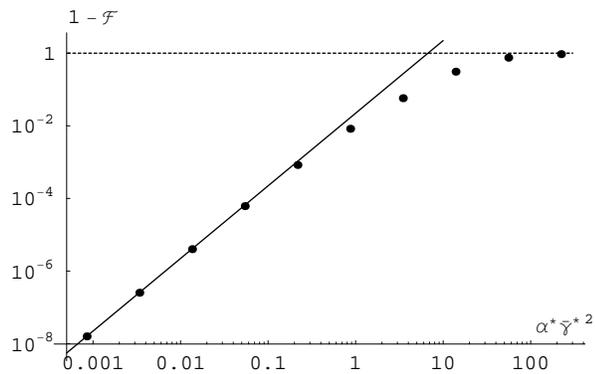}
\caption{The infidelity, $1-\mathcal{F}$ versus non-dimensional diffusion parameter $\alpha^*\bar\gamma^{*\,2}$.  Points are from numerical calculation using a cavity to simulate a dispersive medium.  Solid line is the analytic result, taken from \cite{stace:126804}.  When the dispersion becomes dominant, the infidelity (i.e.\ error) asymptotes to unity.}
\label{infidelity}
\end{center}
\end{figure}

Recall that the conditions for the cavity to accurately simulate weak dispersion are $\Delta^2 = {\sqrt{3}u^3}/{8\alpha L}$ and $\gamma^2=12\Delta^2$, so we solve the master equation, \eqn{cascadeME}, using these parameters.  In order to analyse the dependence of the infidelity, given by $\infid=1-\mathcal{F}$ where $\mathcal{F}$ is the fidelity of the transfer, as a function dispersion, we nondimensionalise the parameters thus: $\alpha^*=\alpha/L u$, $\Delta^*
=\Delta L/u$, $\gamma^*=\gamma L/u$.  In \fig{infidelity} we compare the results of numerical simulations with the heuristic analytic expression given in \cite{stace:126804}.  In that work it was found that the infidelity due to dispersion is given by 
\beq
\infid=(\alpha^*\bar\gamma^{*\,2})^2/45,
\eeq
 in the weakly dispersive limit, $\alpha^*\bar\gamma^{*\,2}\ll1$.  In this regime, both approaches are valid and there is  very good agreement, lending credibility to both.  But our new method shows significant deviation from the approximate result even for $\alpha^*\bar{\gamma}^{*\,2} \gtrsim 1$, for which $\bar{\cal F}$ is still small (of order $10^{-2}$). This regime is at the limit of validity of our approach, according to \erf{lov}, if we say $\delta\omega \sim \gamma$. 

\section{Conclusion}

In this paper we have presented a numerical method for modeling the effect of dispersion in quantum channels connecting a source system to a target system. The method is approximate, and can treat dispersion that is not too strong.  We have also shown how to extend the approach to treat fermionic systems with large flux.  Applying our method to a simple example, for which there existed a previous \emph{ad hoc} analytical result,  showed good agreement between the two methods. For more complicated scenarios, analytical approaches are unlikely to be possible, and our technique may be the only practical approach. 


\begin{thebibliography}{11}
\expandafter\ifx\csname natexlab\endcsname\relax\def\natexlab#1{#1}\fi
\expandafter\ifx\csname bibnamefont\endcsname\relax
  \def\bibnamefont#1{#1}\fi
\expandafter\ifx\csname bibfnamefont\endcsname\relax
  \def\bibfnamefont#1{#1}\fi
\expandafter\ifx\csname citenamefont\endcsname\relax
  \def\citenamefont#1{#1}\fi
\expandafter\ifx\csname url\endcsname\relax
  \def\url#1{\texttt{#1}}\fi
\expandafter\ifx\csname urlprefix\endcsname\relax\def\urlprefix{URL }\fi
\providecommand{\bibinfo}[2]{#2}
\providecommand{\eprint}[2][]{\url{#2}}

\bibitem[{\citenamefont{Gardiner and Zoller}(2000)}]{gar00}
\bibinfo{author}{\bibfnamefont{C.~W.} \bibnamefont{Gardiner}} \bibnamefont{and}
  \bibinfo{author}{\bibfnamefont{P.}~\bibnamefont{Zoller}},
  \emph{\bibinfo{title}{Quantum Noise}} (\bibinfo{publisher}{Springer},
  \bibinfo{year}{2000}).

\bibitem[{\citenamefont{Ji et~al.}(2003)\citenamefont{Ji, Chung, Sprinzak,
  Heiblum, Mahalu, and Shtrikman}}]{ji03}
\bibinfo{author}{\bibfnamefont{Y.}~\bibnamefont{Ji}},
  \bibinfo{author}{\bibfnamefont{Y.}~\bibnamefont{Chung}},
  \bibinfo{author}{\bibfnamefont{D.}~\bibnamefont{Sprinzak}},
  \bibinfo{author}{\bibfnamefont{M.}~\bibnamefont{Heiblum}},
  \bibinfo{author}{\bibfnamefont{D.}~\bibnamefont{Mahalu}}, \bibnamefont{and}
  \bibinfo{author}{\bibfnamefont{H.}~\bibnamefont{Shtrikman}},
  \bibinfo{journal}{Nature} \textbf{\bibinfo{volume}{422}},
  \bibinfo{pages}{415} (\bibinfo{year}{2003}).

\bibitem[{\citenamefont{Chung et~al.}(2005)\citenamefont{Chung, Samuelsson, and
  Buttiker}}]{Chung05}
\bibinfo{author}{\bibfnamefont{V.~S.~W.} \bibnamefont{Chung}},
  \bibinfo{author}{\bibfnamefont{P.}~\bibnamefont{Samuelsson}},
  \bibnamefont{and} \bibinfo{author}{\bibfnamefont{M.}~\bibnamefont{Buttiker}},
  \bibinfo{journal}{cond-mat/0505511}  (\bibinfo{year}{2005}).

\bibitem[{\citenamefont{Stace et~al.}(2004)\citenamefont{Stace, Barnes, and
  Milburn}}]{stace:126804}
\bibinfo{author}{\bibfnamefont{T.~M.} \bibnamefont{Stace}},
  \bibinfo{author}{\bibfnamefont{C.~H.~W.} \bibnamefont{Barnes}},
  \bibnamefont{and} \bibinfo{author}{\bibfnamefont{G.~J.}
  \bibnamefont{Milburn}}, \bibinfo{journal}{Phys. Rev. Lett.}
  \textbf{\bibinfo{volume}{93}}, \bibinfo{eid}{126804} (\bibinfo{year}{2004}).

\bibitem[{\citenamefont{Cirac et~al.}(1997)\citenamefont{Cirac, Zoller, Kimble,
  and Mabuchi}}]{cir97}
\bibinfo{author}{\bibfnamefont{J.~I.} \bibnamefont{Cirac}},
  \bibinfo{author}{\bibfnamefont{P.}~\bibnamefont{Zoller}},
  \bibinfo{author}{\bibfnamefont{H.~J.} \bibnamefont{Kimble}},
  \bibnamefont{and} \bibinfo{author}{\bibfnamefont{H.}~\bibnamefont{Mabuchi}},
  \bibinfo{journal}{Phys.\ Rev.\ Lett.} \textbf{\bibinfo{volume}{78}},
  \bibinfo{pages}{3221} (\bibinfo{year}{1997}).

\bibitem[{\citenamefont{Wiseman}(1997)}]{Wis97}
\bibinfo{author}{\bibfnamefont{H.~M.} \bibnamefont{Wiseman}},
  \bibinfo{journal}{Phys. Rev. A} \textbf{\bibinfo{volume}{56}},
  \bibinfo{pages}{2068} (\bibinfo{year}{1997}).

\bibitem[{\citenamefont{Walls and Milburn}(1994)}]{wal94}
\bibinfo{author}{\bibfnamefont{D.~F.} \bibnamefont{Walls}} \bibnamefont{and}
  \bibinfo{author}{\bibfnamefont{G.~J.} \bibnamefont{Milburn}},
  \emph{\bibinfo{title}{Quantum Optics}} (\bibinfo{publisher}{Springer-Verlag},
  \bibinfo{year}{1994}).

\bibitem[{\citenamefont{Wiseman}(1994)}]{WisemanThesis}
\bibinfo{author}{\bibfnamefont{H.~M.} \bibnamefont{Wiseman}}, Ph.D. thesis,
  \bibinfo{school}{University of Queensland} (\bibinfo{year}{1994}).

\bibitem[{\citenamefont{Carmichael}(1993)}]{Car93a}
\bibinfo{author}{\bibfnamefont{H.~J.} \bibnamefont{Carmichael}},
  \bibinfo{journal}{Phys. Rev. Lett.} \textbf{\bibinfo{volume}{70}},
  \bibinfo{pages}{2273} (\bibinfo{year}{1993}).

\bibitem[{\citenamefont{Gardiner}(1993)}]{Gar93}
\bibinfo{author}{\bibfnamefont{C.~W.} \bibnamefont{Gardiner}},
  \bibinfo{journal}{Phys. Rev. Lett.} \textbf{\bibinfo{volume}{70}},
  \bibinfo{pages}{2269} (\bibinfo{year}{1993}).

\bibitem[{\citenamefont{Stace and Barnes}(2002)}]{sta02}
\bibinfo{author}{\bibfnamefont{T.~M.} \bibnamefont{Stace}} \bibnamefont{and}
  \bibinfo{author}{\bibfnamefont{C.~H.~W.} \bibnamefont{Barnes}},
  \bibinfo{journal}{Phys. Rev. A} \textbf{\bibinfo{volume}{65}},
  \bibinfo{pages}{062308} (\bibinfo{year}{2002}).

\end{thebibliography}

\end{document}